\documentclass[12pt,a4paper]{article}

\usepackage{latexsym}
\usepackage{amsmath}
\usepackage{amsfonts}
\usepackage{amssymb}
\usepackage{amsthm}
\usepackage{amscd}
\usepackage{mathrsfs}

\newcommand{\be}{\begin{equation}}
\newcommand{\ee}{\end{equation}}
\newcommand{\bea}{\begin{eqnarray}}
\newcommand{\eea}{\end{eqnarray}}

\def\cP{{\cal P}_y}                     %
\def\cG{{\cal G}}                       %
\def\G{{\cal G}}                        %
\def\cH{{\cal H}}                       %
\def\cI{{\cal I}}                       %
\def\ri{{\mathrm{i}}}                   %
\def\cJ{{\cal J}}                       %
\def\bR{{\mathbb R}}                    %
\def\bC{{\mathbb C}}                    %
\def\bZ{{\mathbb Z}}                    %
\def\bT{{\mathbb T}}                    %
\def\1{{\mbox{\boldmath $1$}}}          %
\def\tr{\mathrm{tr\,}}                  %
\def\jp{\frac{1}{2}}                    %
\def\om{\omega}                         %
\def\omfs{ \omega_{\mathrm{FS}}}        %
\def\val{\vert  y \vert}                %
\def\cp{\bC P(n-1)}                     %
\def\cE{{\cal E}}                       %
\def\cD{{\cal D}}                       %
\def\Hb{H^\mathrm{loc}}                 %
\def\fR{{\mathfrak{R}}}                 %
\def\fS{{\mathfrak{S}}}                 %
\def\cC{{\cal C}}                       %
\def\IIIb{\mathrm{III}_\mathrm{b}}      %
\def\cZ{{\mathcal Z}}                   %
\def\loc{\mathrm{loc}}                  %
\begin{document}

\vspace*{0.5cm}
\begin{center}
{\Large \bf The Ruijsenaars self-duality map as a mapping class symplectomorphism}
\end{center}

\vspace{0.2cm}

\begin{center}
L. Feh\'er${}^{a}$ and C. Klim\v c\'\i k${}^b$  \\

\bigskip

${}^a$Department of Theoretical Physics, WIGNER RCP, RMKI \\
H-1525 Budapest, P.O.B.~49,  Hungary, and\\
Department of Theoretical Physics, University of Szeged\\
Tisza Lajos krt 84-86, H-6720 Szeged, Hungary\\
e-mail: lfeher@rmki.kfki.hu

\bigskip

${}^b$Institut de math\'ematiques de Luminy
 \\ 163, Avenue de Luminy \\ F-13288 Marseille, France\\
 e-mail: klimcik@iml.univ-mrs.fr

\bigskip

\end{center}

\vspace{0.2cm}

\begin{abstract}
This is a brief review\footnote{To appear in the proceedings of
``Lie Theory and its Applications in Physics IX'' (Varna, June 2011).}
of the main results of our paper arXiv:1101.1759  that contains a complete global
treatment of the compactified trigonometric
Ruijsenaars-Schneider system by quasi-Hamiltonian reduction. Confirming
previous conjectures of Gorsky and collaborators,
we have rigorously established the
interpretation of the system  in terms of flat $SU(n)$ connections
on the one-holed torus and demonstrated  that its  self-duality symplectomorphism represents
the natural action of the standard mapping class generator $S$ on the phase space.
The pertinent quasi-Hamiltonian reduced  phase space turned out to be
symplectomorphic to the complex projective space
equipped with a multiple of the Fubini-Study symplectic form and two toric moment maps
playing the roles of particle-positions and action-variables that are exchanged by the duality map.
Open problems and possible directions for future work are also discussed.
\end{abstract}

\newpage
\section{Introduction}

In his study of action-angle maps, Ruijsenaars \cite{SR-CMP} discovered an
intriguing duality relation for
both non-relativistic and relativistic Calogero type
 classical
many-body systems associated to
$A_n$ root systems and rational, hyperbolic or trigonometric interaction
potentials.  In this
   paper our concern is a particular system of that kind, locally
given by the trigonometric Hamiltonian (\ref{3.1}) later on,
which was invented and proved to be self-dual in \cite{RIMS95}.
Our principal goal is to give a self-contained but concise
presentation of the main results
of our detailed work \cite{FK}, where we showed that the global variant
of this system (called
compactified trigonometric Ruijsenaars-Schneider $\IIIb$ system) and its
self-duality
can be naturally understood by means of quasi-Hamiltonian reduction. This
connects the system to the
$SU(n)$ Chern-Simons theory on the one-holed torus, with a special
boundary condition, and
traces back its self-duality symplectomorphism to the standard duality generator
$S\in  SL(2, \bZ)$
of the mapping class group of the one-holed torus. Our results thus
provide rigorous justification of
conjectures put forward over a decade ago by Gorsky and his
collaborators \cite{GN,JHEP} about
the $\IIIb$ system.

The plan of this contribution is as follows.
 In Section 2 we start
with the definition of the concept
of ``Ruijsenaars duality''. In particular, we shall discuss two
alternative, equivalent  definitions of self-duality.
Necessary background information from
quasi-Hamiltonian geometry is summarized next in  Subsection 3.1,
focusing on the example of the internally fused double that will be used subsequently.
Then in Subsection 3.2 we explain how the mapping class group $SL(2,\bZ)$ acts  on every reduced phase space
arising from the double. Section 4 is devoted to expounding the definition of the compactified
$\IIIb$ system.
The main results of \cite{FK}
 are presented in Section 5.
 The content of Section 5 and related further results are discussed
  in Section 6 together with an exposition of  open problems.

\section{The concept of Ruijsenaars duality}

This concept is relevant for classical integrable many-body systems of ``particles'' moving in
1-dimension.
Due to their physical interpretation and Liouville integrability,
these systems  possess ``particle-positions'' and ``action-variables''
that span two Abelian subalgebras in the Poisson algebra of observables.
By definition, two such systems are  in duality
if there exists a symplectomorphism
between their phase spaces that converts the particle-positions of system (i) into
the action-variables of system (ii) and converts the action-variables of system (i) into
the particle-positions of system (ii).
In particular, one speaks of self-duality if the leading Hamiltonians of both systems
(which underlie the many-body interpretation) have the same form.
An alternative  second definition of self-duality is to consider a single
integrable many-body Hamiltonian system $(M,\Omega,H)$, and call it self-dual if there exists
a symplectomorphism $\fS$
of the phase space $(M,\Omega)$ that converts the particle-positions
into the action-variables and the action-variables into the particle-positions.
Notice that the second definition is a special case of the first definition where
the two systems in duality are two copies of the same system and their duality relation is provided by
$\fS$.

If not clear from the context,
we propose the full name of the above duality  be ``Ruijsenaars duality'' or
``duality in the sense of Ruijsenaars'' (also known as action-angle duality).

Let us further discuss the relation between the above two definitions of (Ruijsenaars) self-duality.
To do this, denote by  $\cJ_k$ and $\cI_k$  $(k=1,...,N)$ the particle-positions
and  action-variables for the
system $(M,\Omega, H)$. It is required that there exists  a dense open
submanifold
$M^\loc \subseteq M$ where the symplectic form $\Omega$ is equal to
$\Omega^\loc = \sum_{k=1}^N d\theta_k \wedge d\! \cJ_k$,
with conjugates $\theta_k$
of the $\cJ_k$.
We can view $( \cJ,\theta )$ and $\cI$ as maps from $M^\loc$ into
$\bR^{2N}$ and $\bR^N$, and then have
\be
H^\loc = \cH \circ (\cJ,\theta) = h \circ \cI
\ee
 with some functions
$\cH$ and $h$, where
the form of $\cH$ underlies the many-body interpretation.
Any global symplectomorphism $\fS$  takes $H$ into the integrable Hamiltonian
$\tilde H:= H\circ \fS$.  One has the relations
$\Omega^\loc= \sum_{k=1}^N d\tilde \theta_k \wedge d \tilde \cJ_k$ and
\be
\tilde H^\loc = H^\loc \circ \fS = \cH \circ (\tilde \cJ, \tilde \theta)= h \circ \tilde \cI
\ee
with
$(\tilde \cJ, \tilde \theta ):= (\cJ, \theta) \circ \fS$ and $\tilde \cI:= \cI \circ \fS$.
Thus  $\tilde H^\loc$ has the same form in terms of the tilded-variables as $H^\loc$ in terms
of the  tilde-free  variables.
Now observe that
 the system $(M,\Omega,H)$ is in duality with $(M,\Omega, \tilde H)$
if $\tilde \cJ$ is the same as $\cI$ and $\tilde \cI$ is the same as $\cJ$.
Spelling this out in more detail: if $(M,\Omega,  H)$
is self-dual in the sense of the second definition, then
its dual pair  $(M,\Omega,\tilde H)$ is automatically manufactured and these two systems are
in duality with respect to the
identity map\footnote{
In general, identifying the phase spaces of any dual pair by the  symplectomorphism that appears
in the definition of the duality relation given at the beginning, one may always turn
this symplectomorphism into the identity map.
Thus the phase spaces of the systems in duality become models of a single phase space,
(not accidentally) similar to two gauge slices serving as models of the single space of gauge
orbits in a gauge theory.}
on $M$.
The full equivalence of our alternative definitions of self-duality is also not difficult to prove.
In this paper we adopt the second definition.

To be precise, we note that in the statement ``is the same as'' above one
must admit some some sign change
or re-labeling of the indices of the  variables.
In fact, the \emph{self-duality symplectomorphism} $\fS$ is usually  not
an involution but has order $4$.
As an illustration, consider the free system with  Hamiltonian
$H= p^2$ on the phase space $\bR^2 = \{ (q,p)\}$, whose particle-position and action-variable
are  $q$ and $p$, respectively.
The free system is trivially self-dual with
self-duality symplectomorphism $\fS: (q, p) \mapsto (p,-q)$,
and dual Hamiltonian $\tilde H= q^2$.

Ruijsenaars \cite{SR-CMP,RIMS95} actually found three distinct
dual pairs of systems and three  self-dual systems.
For example, the dual of the hyperbolic Sutherland system is the rational Ruijsenaars-Schneider
system, and the rational Calogero system is self-dual. See the review \cite{SR-CRM} for the other cases.
Incidentally, at the quantum mechanical level,
all these systems are known to enjoy the related bispectral property
\cite{DG}, too.
As was already mentioned,
in this paper our concern will be the self-dual $\IIIb$ system.
For a detailed geometric treatment of a very different, not self-dual,
case of the trigonometric Ruijsenaars duality,
the reader may consult \cite{FKinCMP}.

\section{Generalities about the internally fused double $D$}

The basic reference for Subsection 3.1 is \cite{AMM}.
The mapping class group action presented in Subsection 3.2 is also well-known to experts \cite{AMM,Gold, Huebs};
in its explicit description we follow  \cite{FK}.

\subsection{Quasi-Hamiltonian systems on $D$ and their reductions}

Let $G$ be a (connected and simply connected) compact Lie group and
fix a positive definite invariant scalar product $\langle\ ,\ \rangle$ on
its Lie algebra $\cG$.
Equip the Cartesian product
\be
D:=  G \times  G =\{ (A,B)\,\vert\, A,B\in   G\}
\label{2.1}\ee
with the 2-form $\omega$,
\be
 2\omega := \langle A^{-1} dA \stackrel{\wedge}{,} dB B^{-1}\rangle
+\langle  dA A^{-1} \stackrel{\wedge}{,} B^{-1} dB \rangle
-  \langle (AB)^{-1} d (A B) \stackrel{\wedge}{,} (BA)^{-1} d (BA)
\rangle,
\label{2.2}\ee
which is invariant under the $G$-action $\Psi$ on $D$ defined by
\be
\Psi_g: (A,B)\mapsto (  g A  g^{-1}, gB g^{-1}),
\qquad
\forall g\in G.
\label{2.3}\ee
Introduce the $G$-equivariant   map $\mu:D\to G$ by the group commutator
\be
\mu(A,B):= AB A^{-1} B^{-1}.
\label{2.4}\ee
These data satisfy
\be
d\omega =-\frac{1}{12} \mu^*\langle\vartheta,[\vartheta,\vartheta]\rangle,
\quad
\om(\zeta_D,\cdot) = \jp\mu^*\langle\vartheta+\bar\vartheta,\zeta\rangle, \quad \forall \zeta\in\G,
\label{2.5}\ee
\be
\mathrm{Ker}(\om_x)=\{\zeta_D(x)\,\vert\,
\zeta\in \mathrm{Ker}(\mathrm{Ad}_{\mu(x)}+ \operatorname{Id}_\cG)\},
\qquad \forall x\in D,
\label{2.6}\ee
where $\vartheta$ and $\bar\vartheta$ denote,
   respectively, the $\cG$-valued left- and right-invariant Maurer-Cartan forms on $G$
   and $\zeta_D$ generates the infinitesimal action of $\zeta\in \cG$ on $D$.
All this means \cite{AMM} that $(D,\om, \mu)$ is a so-called quasi-Hamiltonian $G$-space
with moment map $\mu$.
This quasi-Hamiltonian $G$-space is nicknamed the internally fused double of $G$.

According to the general theory \cite{AMM},
every $G$-invariant function $h\in C^\infty(D)^G$ induces a unique
vector field $v_h$ on $D$  by requiring that $\omega( v_h, \cdot)=dh$ and ${\cal L}_{v_h}\mu=0$.
The vector field $v_h$ is $G$-invariant and its flow preserves $\omega$.
In this way, $(D,\om,\mu, h)$ yields
a quasi-Hamiltonian dynamical system.
Although $(D,\om)$ is not a symplectic manifold,
one can also introduce an honest Poisson bracket on $C^\infty(D)^G$.
Naturally, for $G$-invariant functions $f$ and $h$ the Poisson bracket is furnished by
\be
\{f, h\}:=\omega(v_f,v_h).
\label{2.7}\ee

Generally speaking, quasi-Hamiltonian systems are of interest since they can be reduced
to true Hamiltonian systems by a generalization of the  Marsden-Weinstein symplectic reduction,
and this can give convenient realizations of important Hamiltonian systems.
To specialize to our case, let us choose a moment map value $\mu_0\in G$
and denote its stabilizer with respect to the adjoint action by $G_0$.
Then consider the space of $G_0$-orbits
\be
P(\mu_0):= \mu^{-1}(\mu_0)/G_0,
\label{2.8}\ee
where $\mu^{-1}(\mu_0):=\{   x\in D\,\vert\,  \mu(x)=\mu_0\}$.
Denote by $\iota:\mu^{-1}(\mu_0)\to D$ the tautological injection
and $p: \mu^{-1}(\mu_0)\to P(\mu_0)$ the obvious projection.
Under favourable circumstances (where the meaning of ``favourable'' is the same as for usual
symplectic reduction), there exists a standard Hamiltonian system $(P(\mu_0),\hat\om, \hat h)$ such that
the  \emph{symplectic} form $\hat \om$ and the reduced Hamiltonian $\hat h$ satisfy the relations
\be
p^*\hat\om=\iota^*\om, \quad p^*\hat h=\iota^*h.
\label{2.9}\ee
The Hamiltonian vector field and the flow defined by
$\hat h$ on $P(\mu_0)$ can be obtained by first restricting
the quasi-Hamiltonian
vector field $v_h$ and its flow to $\mu^{-1}(\mu_0)$ and then applying
the projection $p$.
The Poisson brackets on $(P(\mu_0), \hat \om)$ are inherited from
the Poisson brackets (\ref{2.7})  of the $G$-invariant functions like in
usual symplectic reduction.

Of course, the space of orbits $P(\mu_0)$ is not a smooth manifold in general.
However, it always turns out to be a stratified symplectic space \cite{Huebs},
which means that it is a disjoint union of symplectic manifolds of various
dimensions glued together (in a specific manner).

The symplectic spaces obtained from quasi-Hamiltonian reduction
 always  arise also
from usual symplectic reduction of certain infinite-dimensional manifolds with respect to
infinite-dimensional symmetry groups \cite{AMM}.
In particular, let $\Sigma$ denote the torus with a hole (that is, with an open disc removed);
often called the ``one-holed torus''.
It is known that the moduli space (space of gauge equivalence classes)
of flat principal $G$-connections on $\Sigma$ whose holonomy along the boundary of the hole
is constrained to the conjugacy class of $\mu_0$ is a stratified symplectic space,
which can be canonically identified with the quasi-Hamiltonian reduced phase space
$P(\mu_0)$ in (\ref{2.8}).
It is also worth noting that this space supports two natural \emph{Abelian Poisson algebras}.
Namely, for any $\cH\in C^\infty(G)^G$ let $\cH_1$ and $\cH_2$ denote the $G$-invariant functions on
$D$ given by
\be
\cH_1(A,B):= \cH(A)
\quad\hbox{and}\quad
\cH_2(A,B):= \cH(B).
\ee
The two Abelian Poisson algebras on $P(\mu_0)$ are provided by
\be
\cC^a:= \{ \hat \cH_1 \,\vert\, \cH\in C^\infty(G)^G
\},
\quad \cC^b:= \{ \hat \cH_2 \,\vert\, \cH\in C^\infty(G)^G\}.
\label{2.11}\ee
Note also that $D$ itself can be identified as the space of flat connections on $\Sigma$ modulo
the ``based gauge transformations'' defined by maps $\eta\in C^\infty(\Sigma,G)$ for which
$\eta(p_0)=e$ for a fixed point $p_0$ on the boundary of the removed disc.
The matrices $A$ and $B$ represent the holonomies  of the flat connections along the standard generators of
the fundamental group $\pi_1(\Sigma,p_0)$.

\subsection{Symplectic action of the  mapping class group on $P(\mu_0)$}

Let us consider the (orientation-preserving) mapping class group of the one-holed torus,
\be
 {\rm MCG}^+(\Sigma)\equiv\pi_0( {\rm Diff}^+(\Sigma)),
 \label{2.12}\ee
whose elements are equivalence classes of orientation-preserving diffeomorphisms  up to homotopy.
It is known that the mapping class groups acts by structure preserving smooth maps on every reduced phase
space $P(\mu_0)$ (\ref{2.8}), where ``structure preserving'' means symplectomorphism whenever $P(\mu_0)$
is a smooth manifold.
The origin of the mapping class group action is especially clear in the setting of flat connections, where it
arises from the pull-back of the connection 1-forms by diffeomorphisms.
However, it is also possible to directly describe the mapping class group action on $P(\mu_0)$ by taking advantage
of the quasi-Hamiltonian formalism.

For the one-holed torus there exists a (geometrically engendered) isomorphism
\be
{\rm MCG}^+(\Sigma) \simeq SL(2,\bZ).
\label{2.13}\ee
The infinite discrete group $SL(2,\bZ)$ is generated by two elements $S$ and $T$
subject to the relations
\be
S^2=(ST)^3, \quad S^4=1.
\label{2.14}\ee
As concrete matrices, one may take
\be
S=   \begin{bmatrix}
0 &1  \\
-1 & 0
\end{bmatrix}, \qquad
T =  \begin{bmatrix}
1 &0  \\
1 & 1
\end{bmatrix},
\label{2.15}\ee
which actually represent the action of corresponding mapping classes on the standard basis
of the homology group  $H_1(\Sigma;\bZ)\simeq \bZ^2$.
The mapping class of $T$ is known as a Dehn twist and that of $S$ as the standard
orientation-preserving duality generator
``exchanging'' the standard homology cycles.
By arguments detailed in \cite{FK,Gold}, it is natural to associate to $S$ and $T$ the
following diffeomorphisms
$S_D$ and $T_D$ of the double:
\be
S_D(A,B):=(B^{-1}, BAB^{-1}),
\qquad
T_D(A,B):=(AB,B).
\label{2.16}\ee
It is not difficult to check that
\be
S_D^* \om = \om,
\quad
S_D \circ \Psi_g = \Psi_g \circ S_D,
\quad
\mu \circ S_D = \mu,
\label{2.17}\ee
and similar relations hold for $T_D$ as well, i.e., both $S_D$ and $T_D$ are automorphisms
of the internally fused double.
Moreover, one finds that $S_D$ and $T_D$ satisfy
\be
S_D^2=(S_D\circ T_D)^3, \quad S_D^4=Q,
\label{2.18}\ee
where $Q$ is the central element of the group of automorphisms of the double given by
\be
Q(A,B) = \Psi_{\mu(A,B)^{-1}}(A,B).
\label{2.19}\ee
It is an immediate consequence of the above relations that $S_D$ and $T_D$ descend to
maps $S_P$ and $T_P$ on any reduced phase space $P(\mu_0)$ (\ref{2.8}), and these maps
generate an $SL(2,\bZ)$ action on $P(\mu_0)$. Indeed,
$Q$ descends to the trivial identity map $\mathrm{id}_P$ on $P(\mu_0)$, and thus (\ref{2.18})
implies the identities
\be
S_P^2=(S_P\circ T_P)^3, \quad S_P^4=\mathrm{id}_P.
\label{2.20}\ee
The resulting $SL(2,\bZ)$ action preserves the (stratified) symplectic structure on $P(\mu_0)$.

Finally,  consider the action of $S_P$ on the two Abelian Poisson algebras
$\cC^a$ and $\cC^b$ displayed in (\ref{2.11}).
For any $\cH\in C^\infty(G)^G$, define  $\cH^\sharp \in C^\infty(G)^G$   by
\be
\cH^\sharp(g):= \cH(g^{-1}).
\label{2.21}\ee
Then the following identities hold:
\be
\hat \cH_2 \circ S_P = \hat \cH_1
\quad\hbox{and}\quad
\hat \cH_1 \circ S_P= \hat \cH^\sharp_2,
\quad
\forall \cH\in C^\infty(G)^G.
\label{2.22}\ee
In this way, $S_P$ exchanges the elements $\hat \cH_2$ of $\cC^b$ with the elements $\hat \cH_1$  of $\cC^a$.

\section{Compactified Ruijsenaars-Schneider $\IIIb$ system}

In \cite{RIMS95} Ruijsenaars studied, among others,  a particular real form of the complex trigonometric
Ruijsenaars-Schneider system whose Hamiltonian exhibits periodic dependence both on the
particle-positions and on the conjugate momenta.
This system is termed the $\IIIb$ system, where the label ``b'' indicates the bounded nature
of the underlying phase space.
The $\IIIb$ Hamiltonian given by (\ref{3.1}) below is formally integrable since it admits the sufficient number of
constants of motion in involution. However, true integrability holds only after compactifying
the local phase space, whereby the  Hamiltonian flows become complete.
Here, we first summarize the definition of the local $\IIIb$ system and then present its compactification.
Although the content of this section can be found in \cite{RIMS95}, too, for the sake of readability
we display all definitions in a self-contained manner.

The many-body interpretation of the $\IIIb$ system is based on the
Hamiltonian
\be
\Hb_{y}(\delta, \Theta) \equiv \sum_{j=1}^n \cos p_j \prod_{k\neq j}^n
\left[1 - \frac{\sin^2 y}{
\sin^2 (x_j-x_k)} \right]^{\frac{1}{2}},
\label{3.1}\ee
where   $\delta_j= e^{\ri 2 x_j}$ $(j=1,...,n)$
are interpreted as the positions
of $n$ ``particles'' moving on the circle and the canonically conjugate
momenta $p_j$ encode the compact
variables  $\Theta_j = e^{-\ri p_j}$; the index $k$ in  the product runs over $\{1,2,...,n\}\setminus \{j\}$.
The real coupling constant $y$ can be obviously restricted to the range $0<\vert y\vert < \pi/2$
(and its sign is irrelevant).
Then the reality of $\Hb_{y}$ can be ensured
by choosing the variables from a connected open domain where
$ \vert y\vert < \vert x_j - x_k \vert < \pi- \vert y\vert $ holds for all $j\neq k$.
The non-emptiness of such a domain is guaranteed by the requirement
\be
 0< \vert y \vert < \pi/n.
\label{3.2}\ee
We impose the center of mass condition $\prod_{j=1}^n \delta_j = \prod_{j=1}^n \Theta_j =1$, and
parametrize the variables so that the local phase space of the system gets identified with
\be
M_y^{\mathrm{loc}}\equiv  \cP^0 \times \bT_{n-1},
\label{3.3}\ee
where $\bT_{n-1}$ is the $(n-1)$-torus and $\cP^0$ is the interior of the polytope
\be
\cP:=\Bigl\{(\xi_1,...,\xi_{n-1})\in \bR^{n-1}\,\Big\vert\,
  \xi_j \geq \vert y \vert, \,\,\, j=1,...,n-1, \,\,\,
  \sum_{j=1}^{n-1} \xi_j\leq \pi-\val\Bigr\}.
  \label{3.4}\ee
Using the $n\times n$ matrix $E_{j,j}$ having $1$ in the $jj$ position and the identity matrix  $\1_n$,
we introduce
\be
H_k:= E_{k,k} - E_{k+1,k+1},
\quad
\lambda_k:=  \sum_{j=1}^k E_{j,j} - \frac{k}{n} \1_n,
\quad k=1,...,n-1.
\label{3.5}\ee
Then, for  $\xi \in \cP^0$ and
$\tau = (\tau_1,..., \tau_{n-1}) = (e^{\ri \theta_1},...,e^{\ri \theta_{n-1}})\in \bT_{n-1}$,
we define the diagonal $SU(n)$ matrices
\be
\delta(\xi) := \exp\bigl(-2 \ri \sum_{k=1}^{n-1} \xi_k \lambda_k\bigr),
\quad
\Theta(\tau):= \exp\bigl(-\ri \sum_{k=1}^{n-1} \theta_k H_k\bigr).
\label{3.6}\ee
The choice of $\cP^0$ as the domain
of the particle-positions $\xi$ guarantees the positivity of the expressions under the square root in (\ref{3.1}).
In terms of  the variables $(\xi,\tau)\in \cP^0 \times \bT_{n-1}$,
the symplectic form of the system reads
\be
\Omega^{\mathrm{loc}}:= \frac{1}{2} \tr\!\left( \delta^{-1} d
\delta \wedge \Theta^{-1} d\Theta\right) =
\ri \sum_{k=1}^{n-1} d \xi_k \wedge \tau_{k}^{-1} d \tau_k =
\sum_{k=1}^{n-1}  d \theta_k \wedge d\xi_k.
\label{3.7}\ee
Note that for any diagonal matrix $\cD$ (like $\delta, \Theta$ etc), we apply the
notation $\cD=\mathrm{diag}(\cD_1,...,\cD_n)$.

The Hamiltonian (\ref{3.1}) admits $(n-1)$ Poisson commuting constants of motion given
by independent spectral invariants of the following $SU(n)$-valued local Lax matrix:
\be
L^y_{\mathrm{loc}}(\xi,\tau)_{jl}:=
\frac{e^{\ri y} - e^{-\ri y}}{e^{\ri y}\delta_j(\xi) \delta_l(\xi)^{-1} - e^{- \ri y} }
W_j(\xi,y) W_l(\xi,-y) \Theta_l(\tau) \Delta_l(\tau) \Delta_j(\tau)^{-1}.
\label{3.8}\ee
Here we use the positive functions
\be
W_j(\xi,y):= \prod_{k\neq j}^n  \left[ \frac{ e^{\ri y} \delta_j(\xi)  -
e^{-\ri y} \delta_k(\xi) }
{\delta_j(\xi) - \delta_k(\xi)}  \right]^{\frac{1}{2}},
\label{3.9}\ee
and $\Delta(\tau):= \mathrm{diag}(\tau_1,...,\tau_{n-1},1)$.
The Hamiltonian (\ref{3.1}) is recovered from the local Lax matrix as the real part of the trace
\be
\Hb_{y}(\delta(\xi), \Theta(\tau))= \operatorname{Re}\tr\bigl( L^{y}_{\mathrm{loc}}(\xi, \tau)\bigr).
\label{3.10}\ee
Ruijsenaars \cite{RIMS95} realized that the flows of $\Hb_{y}$ and of its commuting family
are not complete on
$M_y^{\mathrm{loc}}$ , and then completed the local phase space in the way described  below.

Let us consider the symplectic manifold $(\cp, \chi_0 \omfs)$, where
\be
\chi_0:= \pi - n \vert y\vert,
\label{3.11}\ee
and $\omfs$ is the standard Fubini-Study symplectic form.
It is convenient to identify the complex projective space $\cp$  as
the factor space $S_{\chi_0}^{2n-1}/U(1)$
with
\be
S_{\chi_0}^{2n-1} = \bigl\{ (u_1,..., u_n)\in \bC^n \,\vert\,
\sum_{k=1}^n \vert u_k \vert^2 =\chi_0\bigr\}.
\label{3.12}\ee
Let $\cp_0$ be the open dense submanifold of $\cp$ where none of the homogeneous coordinates can vanish.
By utilizing the canonical projection $\pi_{\chi_0}: S^{2n-1}_{\chi_0} \to \cp$, we define a  diffeomorphism
$\cE: M_y^{\mathrm{loc}}  \to \bC P(n-1)_0$ by the  formula
\be
\cE(\xi,\tau) := \pi_{\chi_0}(\tau_1 \sqrt{\xi_1 - \vert y \vert},...,
\tau_{n-1} \sqrt{\xi_{n-1} - \vert y \vert}, \sqrt{\xi_n - \vert y\vert})
\label{3.13}\ee
with $\xi_n:= \pi -\sum_{k=1}^{n-1} \xi_k$.
By using that $\pi_{\chi_0}^* (\chi_0 \omfs)=\ri  \sum_{k=1}^n d\bar u_k \wedge du_k$, one sees
that $\cE$ is  a symplectomorphism
\be
\cE^* (\chi_0 \omfs)= \Omega^{\mathrm{loc}}.
\label{3.14}\ee
Thus
we can identify  $(M_y^{\mathrm{loc}},\Omega^{\mathrm{loc}})$ with the
dense open submanifold $\cp_0$ of the compact phase space $(\cp, \chi_0 \omfs)$.
The crucial fact is that, by means of this identification, the local Lax matrix $L^y_{\mathrm{loc}}$
extends to a smooth (even real-analytic) matrix function on $\cp$.
This fact is actually not difficult to verify \cite{RIMS95,FK}.  From now on we denote
the resulting ``global Lax matrix'' as $L^y$.
Since $L^y\in C^\infty(\cp, SU(n))$ satisfies
\be
L^y \circ \cE = L^y_{\mathrm{loc}},
\label{3.15}\ee
it follows that all the smooth spectral invariants of $L^y_{\mathrm{loc}}$ (like the Hamiltonian (\ref{3.10}))
extend to smooth functions on the compactified phase space $\cp$.
The corresponding Hamiltonian flows are automatically complete on $\cp$, simply since every
smooth  vector field has complete flows on a compact manifold.
By definition, the compactified $\IIIb$ system is
the integrable system on the phase space
$(\cp, \chi_0 \omfs)$ whose commuting Hamiltonians are generated by the Lax matrix $L^y$.

\section{Self-duality of the $\IIIb$ system  from reduction}

The compactified  $\IIIb$ system, encapsulated by the triple
\be
(\cp, \chi_0 \omfs, L^y),
\label{4.1}\ee
 possesses two distinguished Abelian Poisson algebras of observables.
The first Abelian algebra is generated by the  ``global particle-position  variables'' $\cJ_k$ defined by
\be
\cJ_k \circ \pi_{\chi_0}(u) = \vert u_k \vert^2 + \vert y\vert,
\qquad
k=1,...,n-1.
\label{4.2}\ee
The terminology is justified by the identity $\cJ_k(\cE(\xi,\tau)) = \xi_k$.
The $\cJ_k$ are the components of the toric moment map
\be
\cJ := (\cJ_1,...,\cJ_{n-1}): \cp \to \bR^{n-1}
\label{4.3}\ee
that generates the so-called rotational action of the torus $\bT_{n-1}$ on $(\cp,\chi_0 \omfs)$.
Its image is the closed polytope $\cP$ (\ref{3.4}).
The other distinguished Abelian algebra is spanned by the action-variables furnished
by certain spectral functions of the global Lax matrix $L^y$.

In the rest of this section we take
\be
G:= SU(n),
\qquad
\langle X, Y\rangle := -\frac{1}{2}\tr(XY),
\quad
\forall X,Y\in \cG.
\label{4.4}\ee
Define the polytope $\mathcal{P}_0$ similarly to (\ref{3.4}) and also define
$\delta(\xi)$ like in (\ref{3.6}) for any $\xi \in \mathcal{P}_0$.
It is well-known that any $g\in G$ is conjugate to a matrix $\delta(\xi)$ for a unique $\xi\in \mathcal{P}_0$,
and $g$ is regular (has $n$ distinct eigenvalues) if and only if the corresponding $\xi$ belongs to the interior
$\mathcal{P}_0^0$ of $\mathcal{P}_0$.
Therefore we can uniquely define a $G$-invariant (i.e.~conjugation invariant) function
 $\Xi_k$ on $G$ by requiring that
\be
\Xi_k(\delta(\xi)) = \xi_k,
\quad \forall \xi \in \mathcal{P}_0,\quad
k=1,...,n-1.
\label{4.5}\ee
The ``spectral function'' $\Xi_k$ is continuous on $G$ and its restriction to
the dense open submanifold of regular elements, $G_{\mathrm{reg}}$, belongs to $C^\infty(G_{\mathrm{reg}})^G$.

It was shown in \cite{RIMS95}, and follows readily from our Theorem 1 given below, that the global Lax matrix
$L^y$ takes values
in $G_{\mathrm{reg}}$ and  the  functions
\be
\cI_k := \Xi_k \circ L^y
\label{4.6}\ee
can serve  as action-variables of the compactified $\IIIb$ system.
In fact, these functions
Poisson commute and their Hamiltonian flows are $2\pi$-periodic.
The image of the toric moment map
\be
\cI:= (\cI_1,...,\cI_{n-1}): \cp \to \bR^{n-1}
\label{4.7}\ee
is the same  polytope $\cP$ as the image of moment map $\cJ$.

One can check that the spectral functions satisfy
\be
\Xi_k^\sharp = \Xi_{n-k},
\label{4.8}\ee
where we applied the definition (\ref{2.21}). Thus, if we define the spectral Hamiltonians
$\alpha_k$ and $\beta_k$ on $D$ by
\be
\alpha_k(A,B):= \Xi_k(A)
\quad\hbox{and}\quad
\beta_k(A,B):= \Xi_k(B),
\label{4.9}\ee
then (\ref{2.16}) implies the identities $\beta_k \circ S_D =\alpha_k$ and $\alpha_k \circ S_D = \beta_{n-k}$.
Although they are not globally $C^\infty$, $\alpha_k$ and $\beta_k$ descend to ``reduced spectral Hamiltonians''
$\hat \alpha_k$ and $\hat \beta_k$ on any reduced phase space $P(\mu_0)$ obtained from the double.
As special cases of (\ref{2.22}),
with the $SL(2,\bZ)$ generator $S_P$ they satisfy
\be
\hat \beta_k \circ S_P= \hat \alpha_k
\quad\hbox{and}\quad
\hat \alpha_k \circ S_P = \hat \beta_{n-k},
\quad
\forall k=1,...,n-1.
\label{4.10}\ee

Having the necessary preliminaries at hand, the principal result of our paper \cite{FK}
can be summarized as follows.

\medskip
\noindent
{\bf Theorem 1.} \emph{For the particular moment map value
\be
\mu_0=  {\rm diag}(e^{2\ri y},...,e^{2\ri y},e^{2(1-n)\ri y}), \qquad
0<\vert y\vert < \pi/n,
\label{4.11}\ee
the ``constraint surface''  $\mu^{-1}(\mu_0)$ lies in $G_{\mathrm{reg}} \times G_{\mathrm{reg}}$ and
the reduced phase space $(P(\mu_0), \hat \om)$ is a smooth manifold symplectomorphic to $(\cp, \chi_0\omfs)$.
The maps
\be
\hat\alpha:= (\hat\alpha_1,..., \hat\alpha_{n-1}): P(\mu_0)\to \bR^{n-1}
\quad\hbox{and}\quad
\hat\beta:(\hat\beta_1,...,\hat\beta_{n-1}): P(\mu_0) \to \bR^{n-1}
\label{4.12}\ee
are toric moment maps generating two effective Hamiltonian actions of $\bT_{n-1}$ on
$(P(\mu_0),\hat\om)$. The images of both $\hat\alpha$ and $\hat \beta$  yield the polytope $\cP$ (\ref{3.4}), and
there exists a symplectomorphism
\be
f_\beta: \cp \to P(\mu_0)
\label{4.13}\ee
that satisfies
\be
 \hat\beta_k \circ f_\beta = \cJ_k
\quad\hbox{and}\quad
\hat \alpha_k \circ f_\beta = \cI_k,
\qquad
\forall k=1,...,n-1.
\label{4.14}\ee}
\medskip

Combining Theorem 1 with the generalities reviewed in Subsection 3.2,
we obtain the following important result.

\medskip
\noindent
{\bf Corollary 1.} \emph{The symplectomorphisms $f_\beta^{-1} \circ S_P \circ f_\beta$ and
$f_\beta^{-1} \circ T_P \circ f_\beta$
generate an $SL(2,\bZ)$ action on the compactified $\IIIb$ phase space $(\cp, \chi_0 \omfs)$.
The mapping class duality symplectomorphism
\be
\fS:=f_\beta^{-1} \circ S_P \circ f_\beta
\label{4.15}\ee
acts by exchanging the particle-positions $\cJ_k$ with the action-variables $\cI_k$ according to
\be
\cJ_k \circ \fS = \cI_k,
\quad\hbox{and}\quad
\cI_k \circ \fS = \cJ_{n-k},
\qquad
\forall k=1,...,n-1.
\label{4.16}\ee}

For the sake of completeness, let us also present the explicit formula
of our map $f_\beta$.
For this, we introduce a unitary matrix $g_y(\xi)$ for each $\xi \in \cP^0$ by
\bea
&&g_y(\xi)_{jn} := - g_y(\xi)_{nj}:= v_j(\xi,y),
\quad
\forall j=1,..., n-1,
\quad
g_y(\xi)_{nn} := v_n(\xi,y),\nonumber\\
&& g_y(\xi)_{jl} := \delta_{jl}- \frac{v_j(\xi,y) v_l(\xi,y)}{ 1 + v_n(\xi,y)},
\quad
\forall j,l=1,..., n-1,
\label{4.17}\eea
where $v_j(\xi,y):=\left[\frac{\sin y}{\sin ny}\right]^{\frac{1}{2}} W_j(\xi, y)$ using (\ref{3.9}).

\medskip
\noindent
{\bf Theorem 2.} \emph{Applying the previous notations, the map $f_0: \cp_0 \to P(\mu_0)$ defined by
\be
(f_0\circ \cE)(\xi,\tau) :=
p \left(g_y(\xi)^{-1} \Delta(\tau) L_{\mathrm{loc}}^y(\xi, \tau)\Delta(\tau)^{-1} g_y(\xi),
g_y(\xi)^{-1} \delta(\xi)g_y(\xi) \right)
\label{4.18}\ee
is a diffeomorphism from $\cp_0$ onto a dense open submanifold of $P(\mu_0)$.
This map is symplectic, $f_0^*\hat\om = \chi_0 \omfs$, and it extends to a global diffeomorphism
$f_\beta: \cp \to P(\mu_0)$.}

\medskip
The map $f_\beta$ that extends $f_0$ automatically has the properties mentioned in the
Theorem above. The statement that $f_0$ is symplectic and that it extends to a global
diffeomorphism  were quite non-trivial to prove.
In \cite{FK}\footnote{The correspondence
$L_{\mathrm{loc}}^y(\xi, \tau)\equiv \Delta(\tau)^{-1} L_y^{\mathrm{loc}}(\delta(\xi),\rho(\tau)^{-1})\Delta(\tau)$
between the respective notations should be noted for those wish to see the details in \cite{FK}.}
the extended map $f_\beta$ was also given explicitly
by making use
of a covering of $\cp$ by $n$ coordinate patches and giving $f_\beta$ explicitly on each patch.

To conclude this section, we remind that
an integrable many-body system
is self-dual in the sense of Ruijsenaars if there exists a symplectomorphism that exchanges its
particle-position variables
with the action-variables.  Hence the message of equation (\ref{4.16}) is that
\emph{our mapping class symplectomorphism $\fS$ (\ref{4.15}) qualifies as
a self-duality symplectomorphism in the sense of Ruijsenaars.}
In fact, we have also checked that $\fS$ coincides precisely with the self-duality symplectomorphism
of the $\IIIb$ system
constructed originally  by a very different (non-geometric, direct) method in \cite{RIMS95}.

\section{Further results and open problems}

This section contains a collection of remarks concerning the results of \cite{FK} and open problems.

First of all, let us recall that every quasi-Hamiltonian reduction of the internally fused double
 represents the
moduli space of flat connections on the one-holed torus $\Sigma$ with fixed conjugacy class of the holonomy around
the hole. This is also the classical phase space of the Chern-Simons field theory on
the three-dimensional  manifold $[0,1] \times \Sigma$ with corresponding boundary condition.
Therefore, our results outlined in the previous section prove the Chern-Simons interpretation
of the $\IIIb$ system and that of its self-duality, confirming
the conjectures of  Gorsky and his collaborators \cite{GN,JHEP}.

In addition to the coupling constant, $y$, a second parameter, $\Lambda$, can be introduced
into the $\mathrm{III}_\mathrm{b}$ system by replacing the symplectic form (\ref{3.7})
by $ \Lambda \Omega^{\mathrm{loc}}$.
This parameter,
which is important at the quantum mechanical level, can be incorporated into the reduction approach by
taking the invariant scalar
product  on $su(n)$ to be  $-\frac{\Lambda}{2}\tr$ instead of (\ref{4.4}).
The quantum mechanics of the
$\IIIb$ system was studied by van Diejen and Vinet \cite{vDV}, who diagonalized the relevant
commuting difference operators using Macdonald polynomials;
see also our note \cite{TMP}  where we reproduced the joint spectrum of the
action-variables by a simple argument.
The Hilbert space of the Chern-Simons theory can be always equipped with a representation
of the mapping class group \cite{wi}, and
it could  be interesting to elaborate this representation in the specific case
 of the $\IIIb$ system by building on the work \cite{vDV}.

Ruijsenaars \cite{RIMS95} also considered an anti-symplectic involution $\fR$
on $\cp$ that enjoys
\be
\cJ_k \circ \fR = \cI_k,
\qquad
\cI_k \circ \fR = \cJ_k,
\qquad
k=1,...,n-1,
\ee
and is given by  $\fR = \hat C \circ \fS$ where $\hat C$ is the complex conjugation involution.
We have shown \cite{FK} that $\fR$ arises from the map $R_D$ of the double of $SU(n)$ defined by
\be
R_D:= \varrho_D \circ S_D^2,
\qquad
\varrho_D(A,B):= (\bar B, \bar A),\qquad \forall (A,B)\in D.
\ee
Although $R_D$ is not quite an automorphism of $D$, it descends to a map $R_P$ on any reduced phase
space $P(\mu_0)$ with diagonal constant
matrix $\mu_0$. (If $\mu_0$ and $\mu_0'$ are conjugate then $P(\mu_0)$ and $P(\mu_0')$ are
naturally equivalent, and therefore one may take $\mu_0$ diagonal without loss of generality.)
The involution $R_P$ reverses the sign of the induced Poisson structure on $P(\mu_0)$,
and together with $S_P$ and $T_P$ it generates a $GL(2,\bZ)$ action on $P(\mu_0)$.

Let $\cal Z$ be the center of the group $G$. Notice that $\cZ \times \cZ$
acts on the internally fused double $D=G\times G$ by the automorphisms
\be
(z_1, z_2): (A,B) \mapsto (z_1 A,z_2 B),
\qquad
\forall (z_1, z_2) \in \cZ \times \cZ.
\ee
This action descends to the reduced phase space $P(\mu_0)$, and in the special case $G=SU(n)$ and $\mu_0$ (\ref{4.11})
it gives rise to the
$\bZ_n \times \bZ_n$ action on $\cp$ used in some considerations in \cite{RIMS95}.

The reader is invited to study \cite{FK} for further results, which include for example
the factorization of $S_D$ as a product of three Dehn twist automorphisms of the double, where the Dehn
twist automorphisms themselves
are realized in terms of certain quasi-Hamiltonian flows.

It could be worthwhile to explore the structure of the stratified symplectic spaces
$P(\mu_0)$ in general, and to possibly uncover new integrable systems on them.
Some sort of trigonometric spin Ruijsenaars-Schneider systems are expected to arise in this way, which might
be integrable analogously to spin Sutherland systems \cite{Resh}.

Finally, the most intriguing open problem stems from the fact that a reduction treatment of the
self-dual hyperbolic Ruijsenaars-Schneider system (the one which is related for example to sine-Gordon solitons)
is still missing.
Presently we do not know what master phase space should give this system upon reduction.
Is it possible to construct such a master phase space?
Of course, there exist other important variants of the Ruijsenaars-Schneider system
($BC_n$ case \cite{Pusztai}, elliptic systems) that should be further studied as well.

\bigskip
\bigskip
\noindent{\bf Acknowledgements.}
We wish to thank the organizers,  V. Dobrev in particular,
for the pleasant atmosphere that we enjoyed at the LT-9 workshop
in Varna.
This work was supported in part by the Hungarian Scientific Research Fund (OTKA) under the grant K 77400.

\end{document}